\documentclass{article}
\pdfoutput=1

\usepackage{arxiv}

\usepackage[utf8]{inputenc} 
\usepackage[T1]{fontenc}    
\usepackage{url}            
\usepackage{booktabs}       
\usepackage{amsfonts}       
\usepackage{nicefrac}       
\usepackage{microtype}      
\usepackage{lipsum}		
\usepackage{amsmath}
\DeclareMathAlphabet{\mathpzc}{OT1}{pzc}{m}{it}

\usepackage{mathrsfs} 
\usepackage{graphicx}
\usepackage{hyperref}       
\usepackage{booktabs}
\usepackage{multirow}
\usepackage[caption=false]{subfig}

\title{Segment Relevance Estimation for Audio Analysis and Weakly-Labelled Classification}

\author{
  Juliano H. Foleiss \thanks{Also as a PhD candidate at the School of Electrical and Computer Engineering, University of Campinas -- Brazil.}\\
  Department of Computing\\
  Federal University of Technology -- Paraná\\
  Campo Mourão, PR -- Brazil \\
  \texttt{julianofoleiss@utfpr.edu.br} \\
   \And
  Tiago F. Tavares \thanks{Also a member of the Interdisciplinary Nucleus for Sound Studies, University of Campinas -- Brazil.} \\
  School of Electrical and Computer Engineering\\
  University of Campinas\\
  Campinas, SP -- Brazil \\
  \texttt{tavares@dca.fee.unicamp.br} \\
}

\begin{document}
\maketitle

\begin{abstract}
We propose a method that quantifies the importance, namely relevance, of audio segments for classification in weakly-labelled problems. It works by drawing information from a set of class-wise one-vs-all classifiers. By selecting the classifiers used in each specific classification problem, the relevance measure adapts to different user-defined viewpoints without requiring additional neural network training. This characteristic allows the relevance measure to highlight audio segments that quickly adapt to user-defined criteria. Such functionality can be used for computer-assisted audio analysis. Also, we propose a neural network architecture, namely RELNET, that leverages the relevance measure for weakly-labelled audio classification problems. RELNET was evaluated in the DCASE2018 dataset and achieved competitive classification results when compared to previous attention-based proposals. 
\end{abstract}


\section{Introduction}

Multiple Instance Learning (MIL) \cite{foulds2010} is a category of learning problems in which labels relate to bags of elements. The bag's label indicate that at least one of its elements is associated with that label. These problems are also known as weakly-labelled problems.
One example of such is the acoustic event detection problem \cite{mesaros19}. In this problem, each audio clip is labelled as a bag, while individual, short-time segments of that clip are its elements. 

Traditional automatic audio classifiers commonly assume that different segments of an audio clip equally contribute towards the sample classification \cite{henaff2011,hamel2011,yandre2017,yu2019}. They implicitly assume that all segments of each audio clip are representative of its label, which does not hold in MIL problems.

This problem was mitigated in previous research by employing attention mechanisms to quantify how much each segment in a clip contributes to the classification. These models have been employed in audio event detection, \cite{kong18,xu2018}, music instrument recognition \cite{gururani2019}, and speech emotion classification \cite{gorrostieta2018,mirsamadi17}.

These attention mechanisms are based on a using a neural network that uses time-domain dependencies in audio signals to quantify the importance of each segment to the overall classification. They assume that the importance of each audio segment to its classification depends only on its content and its time-domain relationship to other segments.

However, this segment importance can depend on the set of classes used in each particular classification problem. For example, identifying if an audio track belongs to the Rock or the Classical genre can rely on several cues, like the presence of distorted guitars, drums, and strong vocals, which are located in many excerpts of Rock tracks. However, classifying a track between Rock and Heavy Metal requires a more careful analysis of the sounds of the instruments, and it is likely that not all excerpts are useful for this classification.

We present an audio classification approach that relies on an importance measure, namely \textit{relevance}, that adapts to the classes regarded in each individual classification problems. Unlike previous approaches, the proposed method does not require re-training when changing the classes for a particular classification problem. Hence, the relevance measure can quickly adapt to different user perspectives, which is useful for exploring diverse viewpoints in computer-assisted audio analysis. Also, it has shown competitive results in the problem of automatic audio event detection.

\section{Proposed Method}
\label{sec:sre_proposed}

Our method uses a set of $N$ one-versus-all classifiers $E_n$, $n \in  \{0, 1, 2...N\}$, which we refer to as \textit{experts}. A different expert is trained for each class in the specific classification problem. This training process uses all elements from the corresponding class as positive segments, and elements from other classes as negative examples. In the prediction process, each expert $E_n$ emits an estimate of the probability $P_n(k)$ that instance $k$ belongs to class $n$. 

If only one expert predicts a high value for $P_n(k)$, then $k$ is associated to a low prediction uncertainty. As a consequence, it can be considered a relevant instance for the bag classification. Conversely, if all experts predict high values for $P_n(k)$, or if all experts predict low values for $P_n(k)$, then $k$ has a high prediction uncertainty and cannot be considered relevant to classify the corresponding bag.

This uncertainty can be measured using Shannon's Entropy:
\begin{equation}
    H(k) = - \sum_{n=1}^{N} P_n(k) \log_n{P_n(k)}.
\end{equation}

Shannon's entropy approaches $1$ when all outcomes are equally probably, meaning that uncertainty about the outcome is maximum. Conversely, it approaches $0$ when one of the events is much more likely than the others, indicating low uncertainty. We use this interpretation to estimate the importance of an element $k$ as the complement of the entropy, that is:
\begin{equation}
    R(k) = 1 - H(k). \label{eq:sre_relevance}
\end{equation}

Equation \ref{eq:sre_relevance} can generate misleadingly high results when $P_n(k)$ is low for all $n$. In this case, the importance of $k$ to the overall result is low even if $R(k)$ is high. This problem can be solved by weighting $R(k)$ by the maximum value of $P_n(k)$ for that bag element, that is,
\begin{equation}
    R_{\text{max}}(k) = R(k)  \max_n {P_n(k)}, \label{eq:sre_relevance2}
\end{equation}
so that high $R_{\text{max}}(k)$ values immediately indicate segments that are useful for classifying the considered bags.

Equation \ref{eq:sre_relevance2} allows adapting the relevance measure to different sets of labels by simply changing the experts used in the classification process. This allows quickly recalculating the importance of particular segments under different perspectives of the same problem, without requiring retraining of the label models. Such a characteristic is useful for computer-assisted audio analysis, as will be discussed in the next section.

\section{Computer-Assisted Audio Analysis}
\label{sec:sre_analysis}
The relevance score presented in Section \ref{sec:sre_proposed} can be used to indicate audio track excerpts that can differentiate it from others. Also, the relevance score changes according to the experts chosen for analysis, which means the relevance score can be adjusted according to the analysis context. Moreover, analysts can add and remove pre-trained experts to obtain different relevance scores, which enables a computer-assisted audio analysis application as follows.

\subsection{Method}
For this application, we encoded audio tracks as  128-bin Mel-Spectrograms. All clips were resampled to 44.1 kHz. The STFT used 2048 samples per frame, with 50\% overlap between frames, with a multiplicative Hanning window. To allow batch training, all spectrograms were zero-padded from both sides in the time axis to size $t$.

\begin{figure}
    \centering
    \includegraphics[scale=0.6]{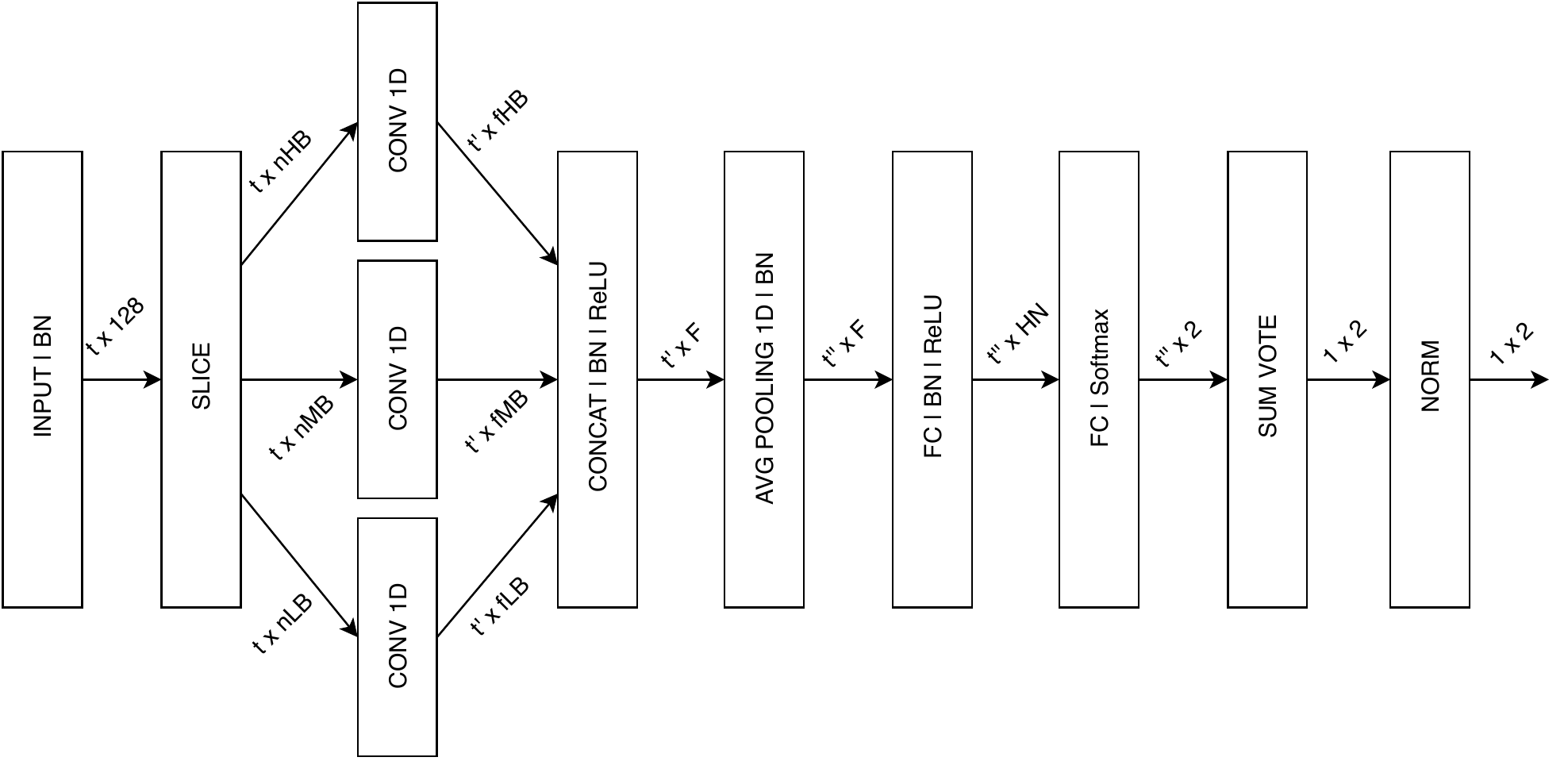}
    \caption{Neural Network Architecture for Expert Classifiers. This network is designed for weak label problems, where labels are given at the clip level. Numbers on the arrows indicate the dimensionality of the output. (Batch dimension is omitted.)}
    \label{fig:sre_expert_arch}
\end{figure}

The experts related to each class were modelled as shown in Figure \ref{fig:sre_expert_arch}. They comprise three parallel 1D-CNNs, each related to a frequency band (low, medium, and high). The bandwise content is obtained by segmenting the Mel-Spectrograms in the frequency axis. The number of frequency bins in each frequency band are referred as \textit{nLB} (low), \textit{nMB} (medium), and \textit{nHB} (high). Each 1D-CNN learns a different number of features, referred to as \textit{fLB} (low), \textit{fMB} (medium), and \textit{fHB} (high).
The outputs of the 1D-CNNs are concatenated into a single vector of size $\text{F} = \text{fLB} + \text{fMB} + \text{fHB}$. These features are aggregated over consecutive segments with average pooling.

The features are the passed on to a fully connected layer with $\text{HN}$ neurons. This layer is followed by a Softmax classifier with 2 outputs, which respectively represent the probabilities that each segment either belongs or does not belong to the expert's class. These estimates are combined by summing the results over all segments, resulting in a $1 \times 2$ vector for each audio excerpt. Last, the outputs are normalised so they sum up to $1$.

Network training used the Adam \cite{kingma2014} optimiser. The loss function was categorical cross-entropy. We avoided over-fitting by using early stopping with patience of $50$ epochs minimal improvement of $0.05$ over the loss. When stopped, the weights from the best epoch were reloaded into the model. The training was done in batches of $100$ audio files, and the maximum number of epochs was set to $1000$.

During training, all bags in the training set are considered positive, while the remaining bags are negative. 20\% of the training instances are used exclusively to evaluate the loss for the early stopping procedure. The test set is entirely separated from the training set.

The relevance score is calculated using the \textit{positive} outputs of the Softmax layer, that is, the estimated probability that each excerpt belongs to the expert's class. These probabilities are used to calculate the relevance using Equation \ref{eq:sre_relevance2}. The relevance score changes according to the chosen experts, that is, it depends on the context.

The expert networks were parameterised as follows: $\text{nLB} = 20$, $\text{nMB} = 40$, $\text{nHB} = 40$, $\text{fLB} = 42$, $\text{fMB} = 42$, $\text{fHB} = 16$, 1D convolution windows over 4 time steps, average pooling window of length 10 with 50\% overlap. This configuration was selected empirically. The number of hidden neurons in the first fully connected layer was evaluated with $\text{HN} = \{20, 50, 100, 150\}$ units.

\subsection{Experiments and Results}
The following experiments aim at demonstrating how changing the regarded classes can modify the relevance score. For such, we trained experts using 90\% of the audio tracks in the GTZAN \cite{tzan2002} dataset. Then, from the remaining 10\%, we picked the track ``Cum on Feel the Noize'', by Quiet Riot, which is labelled as ``Heavy Metal''. It begins with a drum and vocals introduction, and bright, distorted guitars are added close to 15s.

\begin{figure}
    \centering
    \includegraphics[scale=0.48]{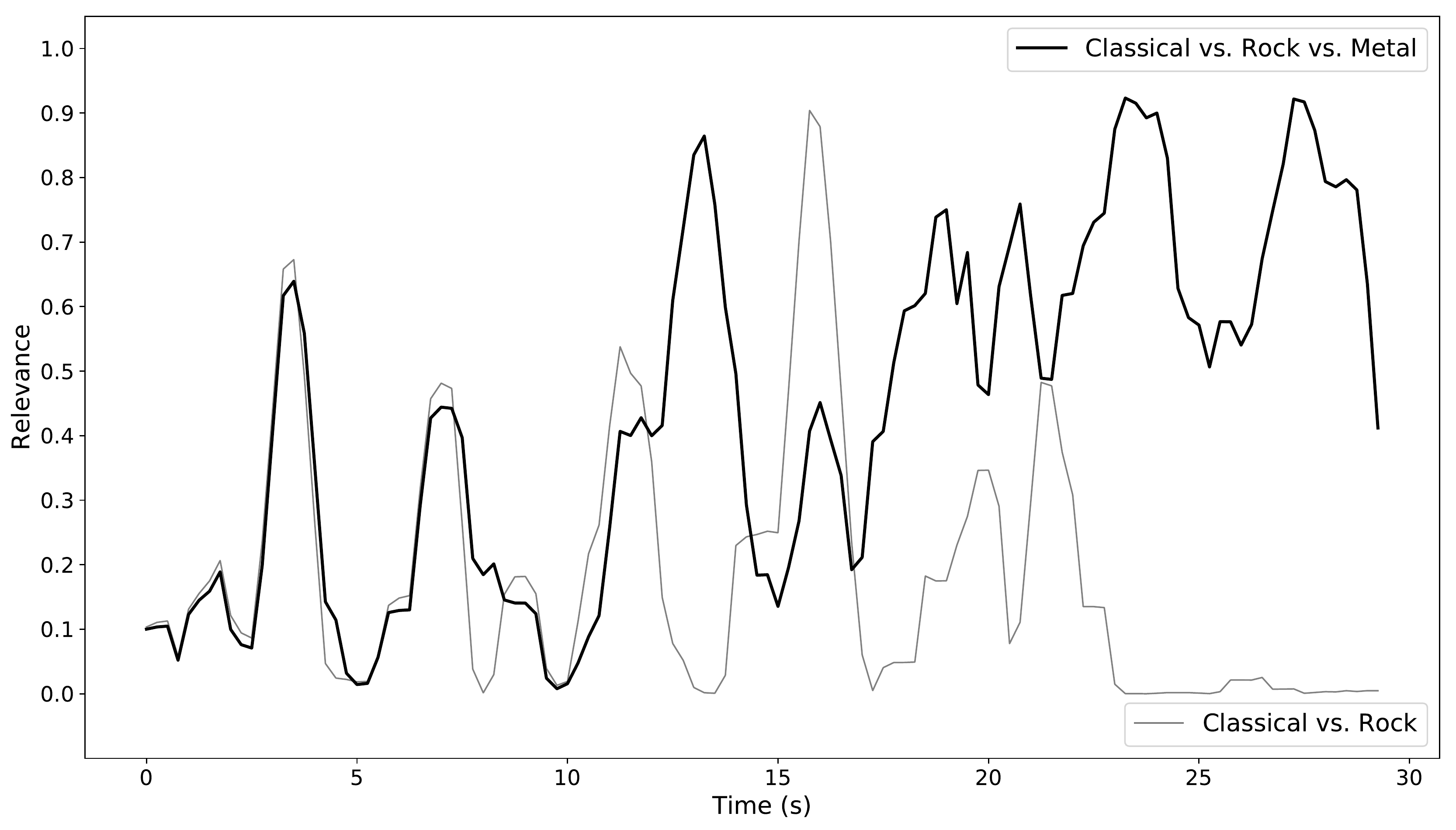}
    \caption{Relevance curves for both Classical vs. Rock and Classical vs. Rock Vs. Metal for ``Cum on Feel the Noize'', by Quiet Riot. The labels on the curves represents the expert that emitted the highest probability overall for each segment.}
    \label{fig:sre_cftn_relevances}
\end{figure}

When only the experts related to ``Rock'' and ``Classical'' genres are used, the relevance, as shown by the light-grey curve in Figure \ref{fig:sre_cftn_relevances}, is higher in the first half of the excerpt, and has a prominent peak close to the drum roll at 15s. This can be viewed as indicative that strong vocals with drive and reverberated drums are seldom used in ``Classical'' music, whereas bright sounds are mildly present both in ``Rock'' and ``Classical'' genres.

If the ``Heavy Metal'' genre is added, the relevance drastically increases around the second half of the excerpt, as shown by the black curve in Figure \ref{fig:sre_cftn_relevances}. This can be related to the fact that distorted bright guitars are typical of the ``Heavy Metal'' genre, whereas drums with strong vocals are more common in ``Rock''.

This illustrates that the importance of each audio segment is inherently linked to the perspective under which classification takes place. The proposed relevance score takes that into account, whereas previous attention mechanisms do not. Next, we show how the relevance can be used in a classification problem.

\section{Acoustic Event Detection}
The problem of Acoustic Event Detection consists of locating specific events within audio signals. The audio signals are labelled according to the contained events, but do not contain the location of the events. In this problem, positive segments (i.e., segments specifically related to the audio event) within bags are sparse. Thus, most audio segments within the recordings are negative instances, hence highlighting the positive segments is important.

\subsection{Method}

\begin{figure}
    \centering
    \includegraphics[scale=0.7]{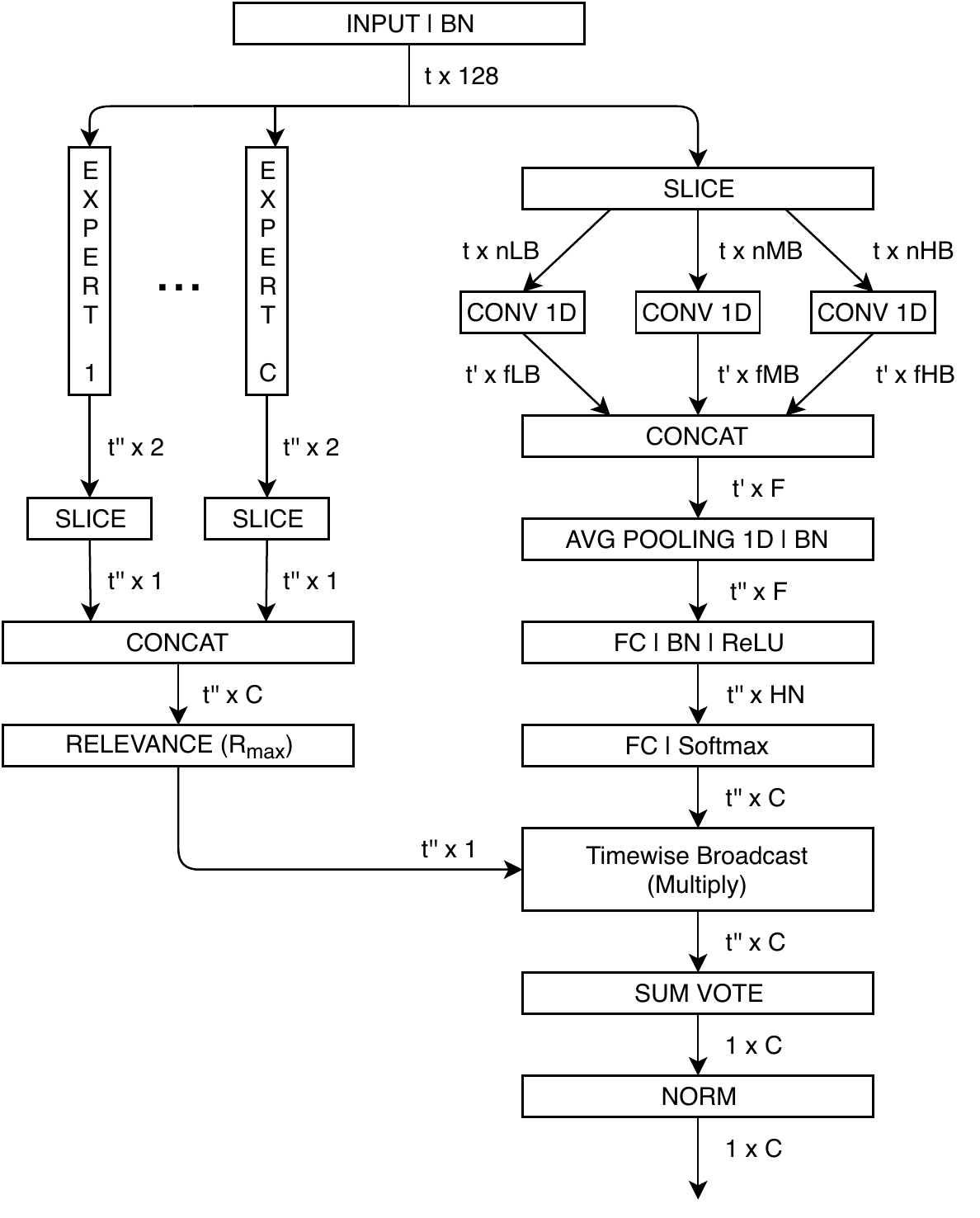}
    \caption{Relevance Network (RELNET). Numbers on the arrows indicate the dimensionality of the output. (Batch dimension is omitted.)}
    \label{fig:sre_relevance_network}
\end{figure}

In all experiments, we used the same audio representation and expert architecture used in Section \ref{sec:sre_analysis}. The full classification network is called \texttt{RELNET}, and is shown in Figure \ref{fig:sre_relevance_network}. This network is comprised of two parallel branches: a relevance network and a classifier network.

The relevance network calculates the relevance score for each audio segment in the bag, while the classifier network predicts the class probabilities for each audio segment. The relevance scores for each segment are calculated using Equation \ref{eq:sre_relevance2}. The output of the relevance network is used as weights for the probabilities output by the classifier network. Finally the last layer aggregates the weighed probabilities by class over all segments, followed by a normalisation to turn it into a probability distribution.

\texttt{RELNET} is trained in two stages. First, all expert classifiers are trained separately as shown in Section \ref{sec:sre_analysis}. In the second stage, the classifier network (as shown in Figure \ref{fig:sre_relevance_network}) is trained. 
During this second stage, the weights of the expert classifiers are not updated. This aims to maintain the interpretability of the relevance branch. 

As a baseline, we used a convolutional neural network (CNN) without any relevance mechanism. For this, we used the network presented in Figure \ref{fig:sre_expert_arch} with the same parameters as the \texttt{RELNET}. We call this architecture \texttt{CONVNET}. This approach is equivalent to using \texttt{RELNET}'s  classification branch alone.

We also evaluated the idea of directly using the experts to output classifications. For such, we used different late fusion procedures: Max Voting (\texttt{MV}), Sum Voting (\texttt{SUM}), Product Voting (\texttt{PROD}), and Relevance Voting (\texttt{RV}), as follows. In \texttt{MV}, the class with maximum probability in each segment gets a full vote, and the classifier decides for the class with the most votes. In \texttt{SUM}, the positive class probabilities for every expert over all segments are aggregated by addition and the classifier decides for class with the highest accumulated probability. \texttt{PROD} is the same as \texttt{SUM}, except that aggregation is done by multiplication. \texttt{RV} is similar to \texttt{MV}, but uses the relevance score (as shown in Equation \ref{eq:sre_relevance2}) as the weight of each vote. \texttt{RV} is equivalent to using \texttt{RELNET}'s  relevance branch alone.

As an attention baseline, we evaluated the attention mechanism in Kong \emph{et al.} \cite{kong18}. In that paper, the authors proposed a non-recurrent, self-attentive mechanism for the acoustic event classification problem. The original model \cite{kong18} was a simple MLP model, and was made available online by the authors. To make a fair comparison, we tweaked their model by changing the first two fully connected layers into the three parallel 1D convolutional layers shown in Figure \ref{fig:sre_expert_arch}. We call this tweaked model \texttt{CONV-KONG}. 

\subsection{Experiments and Results}
We evaluated the relevance-based automatic audio event detector using the DCASE 2018 Task 2 dataset \cite{Fonseca2018_DCASE} (\texttt{DCASE2018T2}). This dataset contains 11073 audio clips (9473 for training and 1600 for testing). All audio clips were automatically labelled, but only about 40\% of the training clips were manually verified, while the remaining were not. All testing clips were manually verified. The labels are a subset of the Audio Set Ontology and consist of 41 classes. 

We zero-padded the inputs so all bags in a batch have the same input dimensions, which allows efficient model fitting using GPUs.
In the \texttt{DCASE2018T2} dataset, around 75\% of the resulting segments are zero vectors after padding.

\subsubsection{Qualitative evaluation}
\begin{figure}
    \centering
    \includegraphics[scale=0.48]{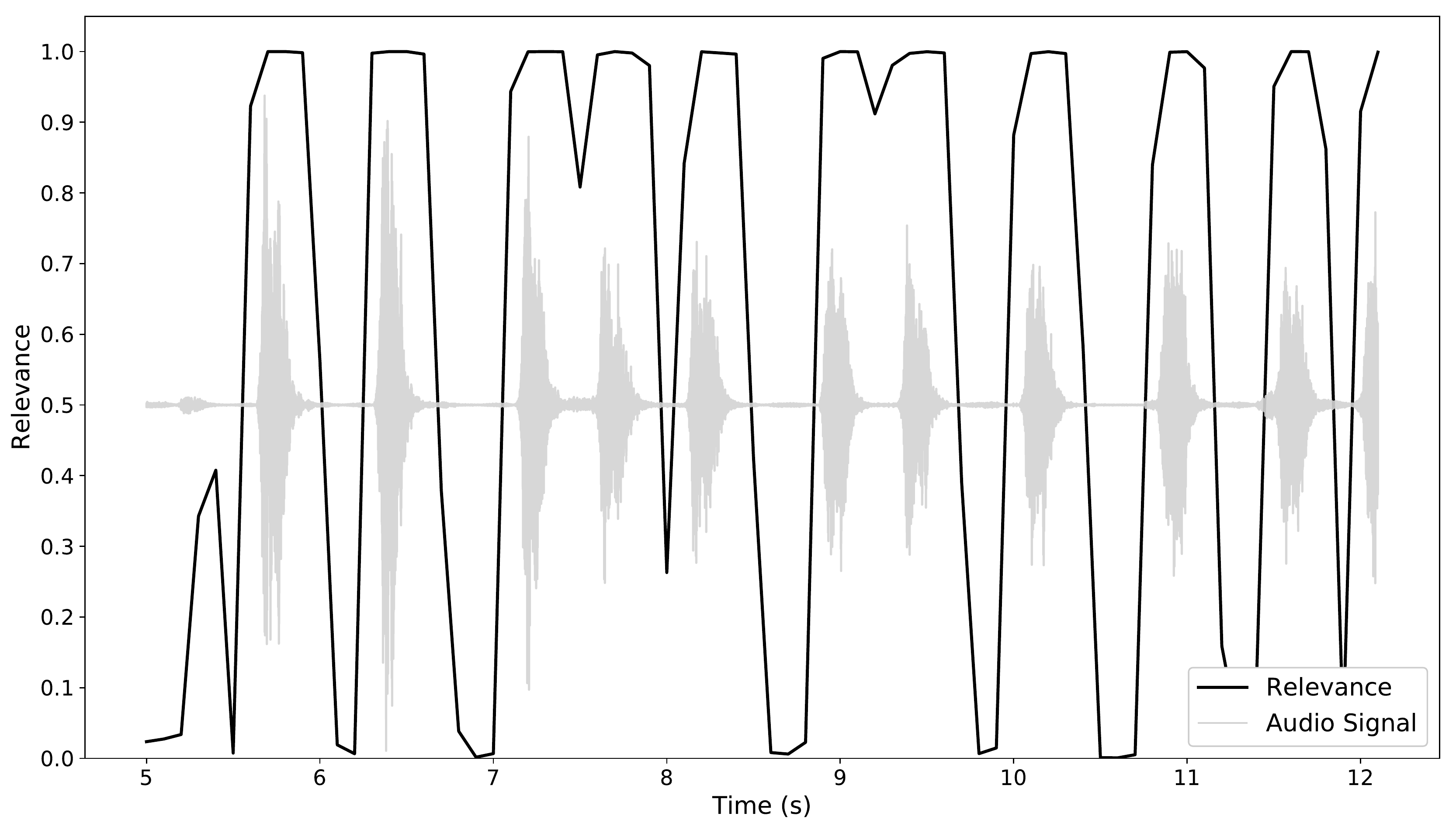}
    \caption{Relevance Plot for 01a5a2a3.wav (Bark). This figure shows the relevance curve from the middle 8s. The corresponding time-domain audio signal is shown in light gray. The following experts were used to calculate the relevance curve: Bark, Bass Drum, Shatter, and Squeak.}
    \label{fig:sre_bark}
\end{figure}

The proposed measure highlights the audio segments that are more relevant for classification. This behaviour is illustrated in Figure \ref{fig:sre_bark}, which shows the relevance plot for an audio clip from the test set of a barking dog (\texttt{01a4a2a3.wav}). The relevance peaks coincide with the dog barking, while the remaining audio segments have low relevance values.

\subsubsection{Quantitative evaluation}
The classification effectiveness was evaluated using the MAP@3 score, as previously used in the DCASE competition \cite{Fonseca2018_DCASE}. This measurement is the number of instances in which the ground-truth label is within the first three predictions of the system.

The results are shown in Table \ref{tab:sre_ae-results}. The left column shows the classification results (MAP@3) when the testing set is also zero-padded to 1200 frames. The right column shows the classification results without padding the test set. 
It is possible to see that, with padding, the results for RELNET, the attention models, and both SUM and RV expert fusions outperformed the CONVNET baseline. Both the CONV-KONG attention model and RELNET achieved better results than the DCASE Competition baseline, which is 0.694 \cite{Fonseca2018_DCASE}.

\begin{table}[h]
\centering
\caption{Classification Results With and Without Zero-padding the Test Set for the DCASE2018 Task 2 Dataset (General-purpose audio tagging of Freesound Content with Audio Set Labels) (MAP@3 Score)}
\label{tab:sre_ae-results}
\begin{tabular}{@{}clcc@{}}
\toprule
\multicolumn{2}{c}{Model}                                                                      & \multicolumn{1}{l}{With Padding} & \multicolumn{1}{l}{W/o padding} \\ \midrule
NN Baseline                                                                     & CONVNET     & 0.590                            & 0.741                           \\ \midrule
This                                                                             & RELNET      & 0.762                            & 0.754                           \\ \midrule
\multirow{2}{*}{Attention}                                                       & KONG \cite{kong18}       & 0.676                            & 0.653                           \\
                                                                                 & CONV-KONG & 0.734                            & 0.649                           \\ \midrule
\multirow{4}{*}{\begin{tabular}[c]{@{}c@{}}Expert\\ Model\\ Fusion\end{tabular}} & MV          & 0.385                            & 0.677                           \\
                                                                                 & SUM         & 0.724                            & 0.719                           \\
                                                                                 & PROD        & 0.040                            & 0.670                           \\
                                                                                 & RV          & 0.716                            & 0.702                           \\ \bottomrule
\end{tabular}
\end{table}

\subsubsection{Discussion}
RELNET clearly outperformed CONVNET when using padding. However, the performance improvement was lower when padding was not used. This indicates that the RELNET architecture allows attenuating the impact of non-informative segments during model fitting.

RELNET also outperformed the expert fusion models. This means that the relevance and classifier branches combined were able to achieve better results than the relevance branch alone.

The relevance values of RELNET can indicate important segments within audio clips, as shown in Figure \ref{fig:sre_bark}. This is similar to the attention mechanism in the KONG and CONV-KONG methods. Our method outperforms both of these methods.

Interestingly, the results obtained by KONG and CONV-KONG are higher (respectively corresponding to a $0.023$ and a $0.085$ improvement) when using padding. This can indicate that the added zeros are being used to perform predictions. Although RELNET has a similar improvement, the performance improvement when using padding is only $0.008$, which is lower than that shown by the baselines.

The results show that the proposed relevance mechanism improves the results in the acoustic events detection problem. Furthermore, it allows pointing the audio segments that are more relevant for the classification process, which facilitates to interpret the results. Next, we show conclusive remarks.

\section{Conclusion}
We present a novel attention-like mechanism, called relevance, which is based on the separability of the target classes in Multiple Instance Learning problems. Our mechanism uses binary classifiers, called experts, to estimate the importance of each segment in an audio clip based on an interpretation of Shannon's Entropy.

The proposed method can be used as a tool for computer-assisted audio analysis. For such, it allows identifying the relevance of audio segments under different perspectives. This can highlight that different parts of an audio excerpt are relevant for classification when the relevant classes change.

Also, relevance can be used to improve classification results in acoustic event detection. For this problem, we proposed a neural network architecture called RELNET that uses the relevance score to highlight the segments in which the events take place. The architecture allows simultaneously improving and interpreting the classification results.

In the future we will explore other applications, such as context-adaptive music indexing, recommendation, and previewing. We will also evaluate our relevance score in other domains, such as in images, text, and video applications.

\bibliographystyle{unsrt}  
\bibliography{tese}  

\end{document}